# Applicability of Well-Established Memristive Models for Simulations of Resistive Switching Devices

E. Linn, *Member, IEEE*, A. Siemon, R. Waser, *Member, IEEE*, and S. Menzel, *Member, IEEE*

*Abstract*—Highly accurate and predictive models of resistive switching devices are needed to enable future memory and logic design. Widely used is the memristive modeling approach considering resistive switches as dynamical systems. Here we introduce three evaluation criteria for memristor models, checking for plausibility of the *I-V* characteristics, the presence of a sufficiently non-linearity of the switching kinetics, and the feasibility of predicting the behavior of two anti-serially connected devices correctly. We analyzed two classes of models: the first class comprises common linear memristor models and the second class widely used non-linear memristive models. The linear memristor models are based on Strukov's initial memristor model extended by different window functions, while the non-linear models include Pickett's physics-based memristor model and models derived thereof. This study reveals lacking predictivity of the first class of models, independent of the applied window function. Only the physics-based model is able to fulfill most of the basic evaluation criteria.

## I. Introduction

Redox-based resistive switches (ReRAM) are an emerging class of two terminal non-volatile devices considered the key components for future memories and logic circuits [1-3]. There are two important subclasses of redox-based bipolar resistive switches. The first one, which is called electrochemical metallization (ECM) or conductive bridge memory (CBRAM), is based on the formation of Cu or Ag filaments [4]. The second one is based on formation of oxygen-deficient filaments in transition metal oxides, so called valence change mechanism (VCM) [5]. Typical VCM devices comprise materials such as $SrTiO_x$, $TaO_x$, $HfO_x$ or $TiO_x$. The field of resistive switching has experienced a substantial progress in 2008 by Strukov et al. suggesting to consider $TiO_x$-based devices as memristive systems [6, 7], or generalized memristors for short [8, 9]. $TiO_x$-based devices are prototypical for the whole class of VCM devices [10], thus modeling all VCM devices as memristive devices should be feasible [9]. The memristive modeling approach enables to capture the complex dynamic behavior of resistive switches by means of a simple ordinary differential equation system which is directly implementable for circuit simulations [6]. The availability of accurate circuit models of resistive switches is the fundament for proper circuit design and further development of novel computer-architectural approaches, as suggested in [11-14], for example.

In general, a memristive system reads

$$\dot{\mathbf{x}} = h(\mathbf{x}, I) \qquad (1)$$

$$V = R(\mathbf{x}, I) \cdot I \qquad (2)$$

where **x** is the inner state variable which is multidimensional in general.

On the basis of Strukov's initial memristor model [7], a whole class of models using different kinds of state variable boundaries, i.e. window functions, has been derived [15-20]. These models are widely used due to simplicity and ease of use [13, 21-22]. In this paper, we evaluate these models with respect to generic properties of resistive switches as well as simulation robustness.

Furthermore, we consider a second class of models which is based on a more complex modeling approach by Pickett et al. and implemented in SPICE by Abdalla et al. [23, 24]. On basis of this physics-oriented modeling approach [25] some simplified and generalized memristive models have been derived [26-29]. We include these models to our comparison and evaluate the properties of these resistive switch models, too.

A comprehensive review of SPICE implementations of several above mentioned models can be found in [30] which was the starting point for our analysis. From the circuit engineer's point of view a model should be as simple as possible, but as complex as required to reproduce essential properties of the device. First of all, to enable a reasonable comparison, we introduce three evaluation criteria extracted from experimental data: 1) the *I-V* characteristics and robustness of the models against parameter input and input signals, 2) the device switching kinetics, and 3) the

E. Linn, A. Siemon and R. Waser are with Institut für Werkstoffe der Elektrotechnik II (IWE II) & JARA-FIT, RWTH Aachen University, Sommerfeldstr. 24, 52074 Aachen, Germany (Corresponding Author e-mail: linn@iwe.rwth-aachen.de)
S. Menzel and R. Waser are with Peter Grünberg Institut 7 (PGI-7) & JARA-FIT, Forschungszentrum Jülich GmbH, 52425 Jülich, Germany

The financial support of the German Research Foundation (DFG) under grant No. LI 2416/1-1 and SFB 917 is gratefully acknowledged.

applicability to simulate complementary resistive switching behavior, i.e. the anti-serial connection of two elements.

## II. EVALUATION CRITERIA

In memristive modeling a resistive switch is considered a dynamical system (equations (1)-(2)). Thus, it should be possible to simulate the device behavior for a wide range of input signals. This is the main strength of the memristive modeling approach compared to using models with built-in fixed threshold voltages [31, 32], which are only valid for certain input signals.

The first evaluation criterion considers the *I-V* characteristics of bipolar resistive switches (Fig. 1a) which exhibit some distinct features that should be reproduced by a suitable model (compare [5], for example).

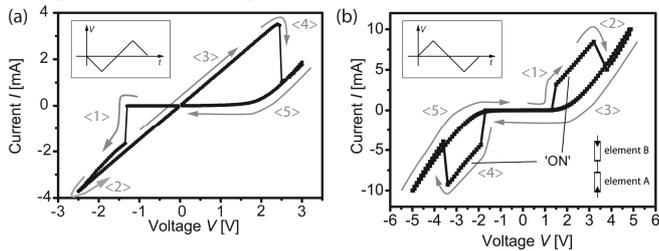

Fig. 1. (a) Exemplary *I-V* characteristic of a TaO$_x$-based resistively switching device. The input signal is a triangular voltage sweep of amplitude -2.5 V/ 3V (see inset). Note that for a symmetric voltage amplitude (e.g., -3 V/ 3V) the characteristic does not change significantly. Initially, the device is in HRS state and switches to state LRS at <1>. In <2>, <3> the device stays in the LRS and switches back to HRS at <4>. This is again the initial HRS state. Note that the switching polarity depends on the actual material composition. Here, the input voltage was applied to the top electrode (Pt) while the bottom electrode (Ta) was grounded. (b) *I-V* characteristic of a TaO$_x$-based complementary resistive switch device. The input signal is a triangular voltage sweep of amplitude 5 V (see inset). Starting from HRS/LRS (element A/element B), the device switches to LRS/LRS ('ON state') first (<1>). Next, the CRS cell switches over to LRS/HRS (<2>) and remains in this state until the negative SET voltage is reached (<3>). Then, the device switches again to LRS/LRS (<4>), and later on back to HRS/LRS (<5>). For details on fabrication see [33].

Typically, during the SET operation an abrupt increase in current is observed for ECM and VCM-based ReRAM cells. The RESET operation, however, differs for these two classes: VCM devices often show a gradual RESET whereas ECM devices exhibit an abrupt change. Furthermore, the *I-V* characteristics are asymmetric with respect to the origin. In addition, the SET and RESET voltages increase when the sweep rate of the voltage sweep is increased. Additionally, we know from experiment that a wide range of excitation signals will lead to resistive switching device behavior. In so far, a suitable memristive model should offer certain robustness against changes in the input voltage amplitude and variations of initial values, e.g. the initial value of the state variable.

The second criterion is related to the switching kinetics. In experiments a strong non-linear relationship between SET time $t_{SET}$ and pulse height $V_p$ is observed (see Fig. 2). Here we selected device data from four typical VCM devices: strontium titanate [34], tantalum oxide [35], hafnium oxide [36], and titanium oxide [37]. A common fingerprint of all VCM devices is the decrease of $t_{SET}$ by several orders of magnitude by only increasing the pulse amplitude $V_p$ by a factor of two. Hence, our second criterion is the check for such an exponential dependency. Fulfilling this criterion is essential to enable simulation of typical applications using memristive devices (either memory or logic applications) which are conducted by fast pulses.

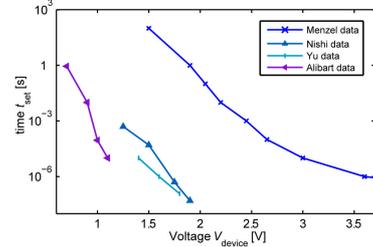

Fig. 2. Set time $t_{SET}$ of the switching from HRS to LRS versus applied pulse height $V_p$. Menzel data for SrTiO$_x$ [34], Nishi data for TaO$_x$ [35], Yu data for HfO$_x$ [36], and Alibart data for TiO$_x$ [37].

A third criterion arises from the need for multi-element simulations for mapping real-world circuits. Two-element circuits are the simplest application, and therefore, are well suited for a basic analysis. Here, we consider an anti-serial connection of two memristive devices, known as complementary resistive switch (CRS) [38]. Typical *I-V* characteristics of VCM-type CRS cells can be found in [33, 39], for example (see Fig. 1b). One distinctive feature of the anti-serial connection is the presence of an overall low resistive state ('ON state') when applying a voltage sweep. The existence of this ON state region in a two element simulation can be used as a further check for model consistency [40].

## III. MEMRISTOR MODELS

### A. Linear model and window function extensions

In [7], the following device equations were suggested:

$$\dot{x} = h(I) = K_1 \cdot I \qquad (3)$$

$$V = R(x) \cdot I = ((R_{LRS} - R_{HRS}) \cdot x + R_{HRS}) \cdot I. \qquad (4)$$

Here, $K_1$ is a constant, $R_{LRS}$ is the resistance of the low resistive state (LRS) and $R_{HRS}$ is the resistance of the high resistive state (HRS). The state variable $x$, which represents the position of the boundary between low conductive and high conductive region, is normalized by the switching layer thickness of $D = 10$ nm.

Note that the equation system (3)-(4) is mathematically not a linear system. However, we call this model a 'linear model' for the following reasons: first, the state variable $x$ influences the resistance $R(x)$ linearly. Second, the voltage $V$ is directly proportional to the current $I$ for a certain $R(x)$. Third, $\dot{x}$ depends linearly on the current $I$. In models from section B, which we call 'non-linear models', at least one of these relations is non-linear.

To prevent nonphysical values for $x$, the state variable must be limited to the layer thickness, thus $0 \le x \le 1$ holds. In common SPICE implementations, this bounding is realized by a window function $f(x,I)$. Thus, equation (3) can be rewritten as:

$$\dot{x} = h(x,I) = K_1 \cdot I \cdot f(x,I). \quad (5)$$

In this study, we consider four different window function implementations: Benderli's model [15], Joglekar's model [16], Biolek's model [17] and Shin's model [18], which are illustrated in Fig. 3 a, c, e and g, respectively. Note that there are several other window function-based models, like Prodromakis's model [19] or Corinto's and Ascoli's model [20, 21] which are not included in this study. We think it will be a worthwhile future task to evaluate these too.

In case of Biolek's model and Shin's model, the window function also depends on the sign of current $I$, compare Fig. 3e, g. However, we can rewrite Biolek's window function as follows:

$$f(x,I) = \begin{cases} 1-(x)^{2p} & \text{for } I \geq 0 \\ 1-(x-1)^{2p} & \text{for } I < 0 \end{cases}. \quad (6)$$

Similarly, Shin's window reads:

$$f(x;I) = \begin{cases} \sigma(1-x) & \text{for } I \geq 0 \\ \sigma(x-1) & \text{for } I < 0 \end{cases}. \quad (7)$$

Here, $\sigma(\cdot)$ is the step function. Thus, when only either purely positive or negative input signals are considered, the window function is only a function of $x$.

The simulation parameters for all models are selected according to [7]: $K_1 = 10^4 \text{ A}^{-1}\text{s}^{-1}$, $R_{LRS} = 100 \text{ }\Omega$ and $R_{HRS} = 16 \text{ k}\Omega$. In Fig. 3c Joglekar's window function is shown. It has the same shape as the Benderli's window function, but can be parameterized with $p$ to vary the gradient. To illustrate the impact of $p$, three curves, for $p = 1$, 7 and 50, are depicted. For the $I$-$V$ simulations in Fig. 3d we applied $p = 1$.

The Biolek's window function offers a similar parameterization as Joglekar's window function, and window curves for $p = 1$, 7 and 50 are shown in Fig. 3e. However, the function behaves quite different due to the involved step function $\sigma(\cdot)$ which enables a sudden upward transition of the window function from a value close to zero towards a value close to unity if the sign of the current $I$ changes and when $x$ is in proximity of its limits (see Fig. 3e). In Fig. 3e the solid line is valid for positive currents (SET direction) and the dashed line for negative currents (RESET direction). Shin's window function (Fig. 3g) can be considered an edge case of Biolek's window function for $p \rightarrow \infty$.

In order to evaluate these models against our first criterion we simulated their $I$-$V$ characteristics. For this, symmetric triangular input voltage signals with sweep rates of 10 V/s, 30 V/s and 100 V/s (see inset in Fig. 3b) are used. The resulting $I$-$V$ characteristics using Benderli's and Joglekar's window function are depicted in Fig. 3b and Fig. 3d, respectively.

Both models exhibit completely symmetric $I$-$V$ characteristic with respect to the origin due to their symmetric window functions, i.e., one switching event (SET) occurs after reaching the maximum voltage level (<2> in Fig. 3b and Fig. 3d) and the other one (RESET) occurs before reaching the maximum absolute voltage level (<4> in Fig. 3b and Fig. 3d). Therefore, the symmetry is an inherent property of these two models and as a consequence they cannot reproduce the asymmetry of bipolar resistive switching, as analytically proved [20]. Keep in mind that Benderli's and Joglekar's model represent memristors with window depending on state only. Fingerprint of these types of models is a symmetric $I$-$V$ characteristic with respect to the origin (compare [20, 41]). Furthermore, Benderli's and Joglekar's model do not show an abrupt SET transition which limits their applicability (cf. Fig. 3b, and d). In contrast, the simulated $I$-$V$ curves using Biolek's and Shin's window function show an abrupt SET transition. In addition, the $I$-$V$ characteristics are asymmetrical with respect to the origin. However, one should note that the characteristics in Fig. 3f, h differ strongly from characteristics of typical VCM devices (see Fig. 1a or [5]) anyway: for example, the current in LRS for negative voltages is very low.

All models offer the general trend of higher SET voltages for increasing sweep rates.

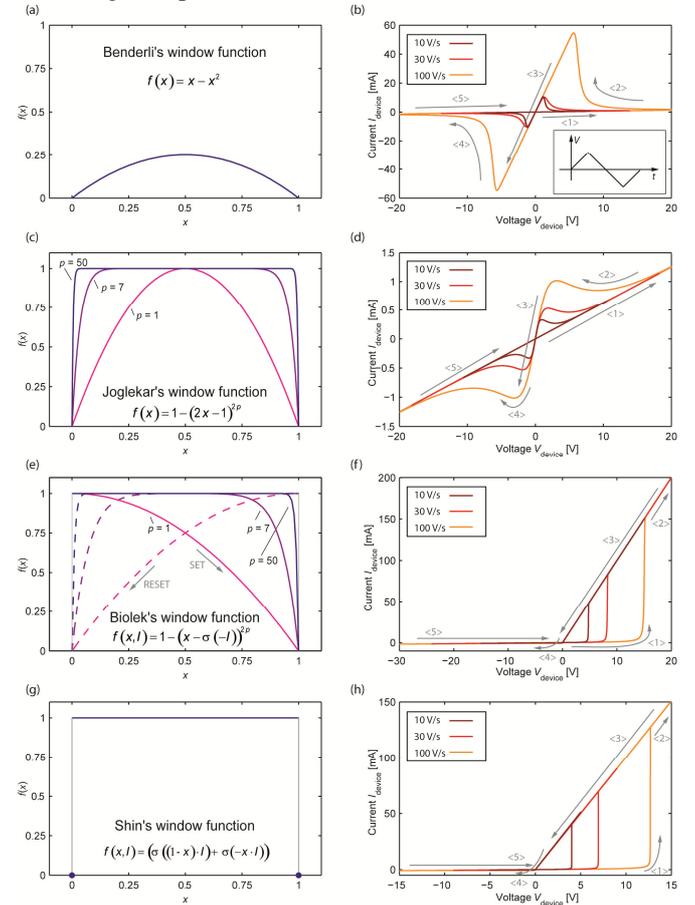

Fig. 3. Implementation of Strukov's model using different window functions. For each model the applied window function and corresponding $I$-$V$-characteristics are depicted. (a,b) Benderli's model, (c,d) Joglekar's model, (e,f) Biolek's model and (g,h) Shin's model. The inset in (b) illustrates the input triangular voltage signal. Arrows and numbers <#> indicate the run of the curve. The initial state $x_0$ for Shin's and Biolek's window was $x_0 = 0$, while for Benderli's window we used $x_0 = 0.002$ and $x_0 = 10^{-12}$ for Joglekar's window.

The window functions offer quite different robustness with respect to input signals. Both the Benderli's model and the Joglekar's model tend to stick at the boundary if voltage amplitudes become larger. This is due to the form of the window function which gives rise to convergence problems (called "terminal-state problem" in [19]), and limits applicability of the model. To enable proper simulation, parameters and input signal must be carefully adjusted for these window functions. This problem was solved in the Biolek and Shin model by resetting the window function when the sign of the input signal changes.

In the following, we evaluate the models with respect to the reproducibility of the experimentally observed switching kinetics, i.e. the second evaluation criterion. Thus, we apply voltage pulses of height $V_{device} = V_p$ to each model being initially in the high resistive state (HRS). For positive voltages the models are SET to the low resistive state (LRS) after the time $t_{SET}$.

From Fig. 4a, we can see that the basic trend observed in experiments, i.e., decreasing $t_{set}$ for increasing pulse height, is also observed for the four simulation models. But, the dependency is much less pronounced as observed in experiments. Compared to experimental data shown in [37] for titanium oxide, we can clearly see that actual dependency differs greatly (several orders of magnitude).

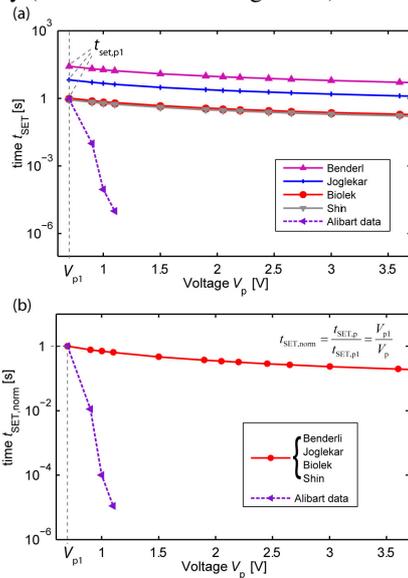

Fig. 4. Set time $t_{SET}$ of the switching from HRS to LRS versus applied pulse height $V_p$. In simulations $t_{SET}$ is defined at $x = 0.5$. (a) shows the raw data. (b) depicts normalized data with respect to the $V_{p1} = 0.7$ V points.

This mismatch can be directly assigned to the $R$-$x$ dependency in model equation (4), which is not sufficiently non-linear. Furthermore, we can show that independent of the applied window function, all models offer the same kinetics (cf. Fig. 4b). By normalizing the $t_{SET}$ values by a certain point (here: $V_{p1} = 0.7$ V), all curves collapse to a single line, showing this feature directly. By inserting equations (4) into equation (5) we obtain:

$$\frac{dx}{dt} = K_1 \cdot \frac{V(t)}{R(x)} \cdot f(x). \tag{8}$$

where we removed the dependence of $f(\cdot)$ on $I$, since for $t \in [t_0, t_1]$ the current has a unique sign and thus $f(\cdot)$ depends only on $x$. Equation (8) is a differential equation offering two variables, $x$ and $t$. Integration by parts results in:

$$\int_{x_0}^{x_1} \left( (R_{LRS} - R_{HRS}) \cdot x + R_{HRS} \right) \frac{1}{K_1 \cdot f(x)} \cdot dx = \int_{t_0}^{t_1} V(t) \cdot dt. \tag{9}$$

The bounds of integration are: $t_0 = 0$ s and $t_1 = t_{set,p}$ for the right side and $x_0 = 0$ and $x_1 = 0.5$ (our assumed SET condition) for the left side of equation (9). Furthermore, we consider a voltage pulse of amplitude $V_p$, thus $V(t) = V_p$ holds. Finally, the equation reads:

$$\int_0^{0.5} \left( (R_{LRS} - R_{HRS}) \cdot x + R_{HRS} \right) \frac{1}{K_1 \cdot f(x)} \cdot dx = V_p \int_0^{t_{SET,p}} dt, \tag{10}$$

which is equivalent to

$$K_2 = V_p \cdot t_{SET,p}. \tag{11}$$

The left hand side of the equation (10) is constant for every model and is called $K_2$ in equation (11). Note that the value of $K_2$ is specific to the applied model, but cancels out when normalizing values with respect to a certain pulse height $V_{p1}$ offering a set time $t_{SET,p1}$. This procedure is done for each model independently, and results in the graph shown in Fig. 4b. Thus, for all models of this kind the normalized SET time only depends on the pulse height $V_p$ and not on the window function (Fig. 4b):

$$t_{SET,norm} = \frac{t_{SET,p}}{t_{SET,p1}} = \frac{V_{p1}}{V_p}. \tag{12}$$

The resulting dependency is

$$t_{SET} \square \frac{1}{V_p}. \tag{13}$$

So, the models are not capable to show the required exponential dependency.

From these considerations it is clear that a simple addition of a window function is not appropriate to introduce realistic device dynamics to the initial memristor model.

Next we consider our third criterion which is the anti-serial connection of two elements. Here, the results reveal even more striking mismatches between simulation and real device behavior. Due to the anti-serial connection of both cells A and B $\dot{x}_A = -\dot{x}_B$ holds if $f(x_A) = f(x_B)$, where we dropped the dependence on $I$. Therefore, any change of state variable in cell A is canceled out by the change of state in cell B. Thus, the total resistance of both elements is constant all the time (see Fig. 5a) which is not the case in reality, as we know from CRS cells (compare Fig. 1b, where $x_{0A} \approx x_{0B} \approx 0$, i.e. $x_A(t=0) \approx 0$ and $x_B(t=0) \approx 1$). For Shin's window function we could show this property already in [42]. Biolek's window (Fig. 3e) shows the same behavior (compare Fig. 5a) while Benderli's window and the Joglekar's window only offer a straight line for symmetrical initial conditions, e.g., $x_{0A} = x_{0B} = 0.001$ ($x_A(t=0) = x_{0A}$ and $x_B(t=0) = 1 - x_{0B}$).

(For the sake of completeness, one must say that for very carefully adjusted input signals and highly asymmetric initial states also Shin and Biolek could show different behavior than Fig. 5a.) For Benderli's window and the Joglekar's window one can force a non-ohmic device behavior by starting from different initial states and considering a smooth window function. For simulation shown in Fig. 5b the model with Joglekar's window ($p = 1$) is used and $x_{0A} > x_{0B}$ is assumed.

Due to the asymmetry of the initial values the SET process in element A starts earlier, leading to a increased current <1> in Fig. 5b. Note that a similar result was observed in [22] which is in accordance to CRS behavior at first glance. However, for the negative voltage cycle the increased current (point <4>) occurs after reaching the maximum absolute voltage which does not correspond to real device behavior at all. For $x_{0A} < x_{0B}$ the observed behavior becomes even more unusual since the resistance (chordal resistance) is increased in a certain regime (grey line curves at points <1>, <4> in Fig. 5c) – the opposite behavior than observed in experiments. In consequence, the simulation result strongly depends on the initial states, which is unfavorable according to the first criterion. However, adjusting the initial states is not suited to reproduce real device behavior for those models. Pay attention that a strong dependency on initial states is a commonly observed incident for memristor models (compare e.g. [14]).

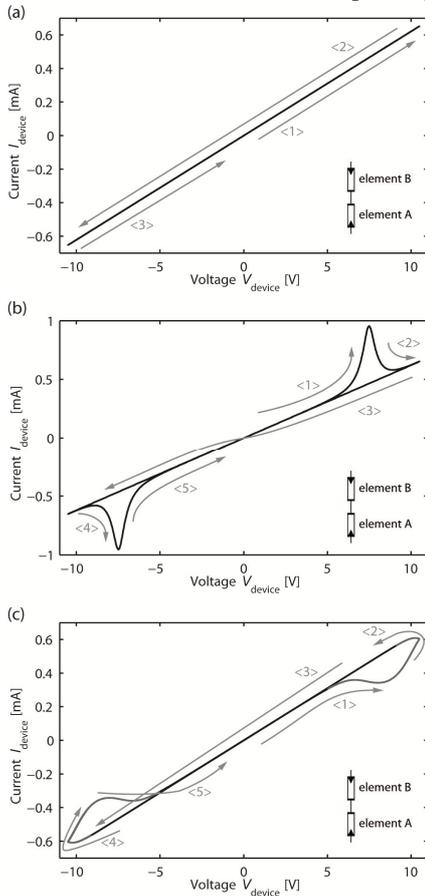

Fig. 5. Anti-serial connected device model simulations using Joglekar's window. A triangular input voltage signal of sweep rates of 10 V/s was used. The curves run-through is denoted by the arrows. (a) For symmetrical initial conditions ($x_{0A} = x_{0B} = 0.001$), no change of the overall resistance is observed. (b) Simulation for $x_{0A} > x_{0B}$ ($x_{0A} = 0.001$ and $x_{0B} = 0.0001$). (c) Simulation for $x_{0A} < x_{0B}$ ($x_{0A} = 0.0001$ and $x_{0B} = 0.001$). The portion of the loop relative to the regime of increased resistance is marked by a grey line color.

In conclusion, although highly attractive due to ease of use, none of the above studied models is suited to reproduce the basic resistive switch properties for arbitrary input signals. But these models could be modified to meet the requirements. The window function can have a positive impact on the accuracy, but cannot fix the basic physical equations. Correspondingly, simulation results obtained from these models, e.g. [12-13, 43], should be reconsidered using more sophisticated models.

However, one should keep in mind that additional model complexity allows a very high predictivity for a variation of parameter inputs. But this complexity might also give rise to convergence issues.

Next, we consider the improved physics-related approach by Pickett [23, 24].

### B. Pickett's model and generalized sinh models

In [23] and [44] an improved modeling approach towards TiO₂ device modeling was suggested and published as a SPICE model by Abdalla and Pickett in [24]. In this model the state variable $w$ corresponds to a tunneling gap, and a highly non-linear current-voltage relationship, i.e., a tunneling current equation, was introduced. By trial and error modification the basic I-V curve could be fitted to a measured I-V curve considering the non-linear switching kinetics of the device. The equations read:

$$I = \frac{J_0 A}{\Delta w^2} \cdot \left\{ \phi_I \cdot \exp\left(-B\sqrt{\phi_I}\right) - \left(\phi_I + e|V_g|\right) \cdot \exp\left(-B\sqrt{\phi_I + e|V_g|}\right) \right\} \quad (14)$$

$$\dot{w} = \begin{cases} f_{\text{off}} \sinh\left(\frac{|I|}{I_{\text{off}}}\right) \exp\left(-\exp\left(\frac{w - a_{\text{off}}}{w_c} - \frac{|I|}{b}\right) - \frac{w}{w_c}\right) & I > 0 \\ -f_{\text{on}} \sinh\left(\frac{|I|}{I_{\text{on}}}\right) \exp\left(-\exp\left(\frac{a_{\text{on}} - w}{w_c} - \frac{|I|}{b}\right) - \frac{w}{w_c}\right) & I < 0 \end{cases} \quad (15)$$

with

$$\phi_I = \phi_0 - e|V_g|\left(\frac{w_1 + w_2}{w}\right) - \left(\frac{1.15 \lambda w}{\Delta w}\right) \ln\left(\frac{w_2(w - w_1)}{w_1(w - w_2)}\right) \quad (16)$$

$$B = \frac{4\pi \Delta w \sqrt{2m}}{h} \quad (17)$$

$$w_2 = w_1 + w\left(1 - \frac{9.2\lambda}{3\phi_0 + 4\lambda - 2e|V_g|}\right) \text{ with } w_1 = \frac{1.2\lambda w}{\phi_0} \quad (18)$$

$$\lambda = \frac{e^2 \ln(2)}{8\pi \kappa \varepsilon_0 w}, \ J_0 = \frac{e}{2\pi h}, \text{ and } \Delta w = w_2 - w_1 \quad (19)$$

$$V_g = V_{\text{device}} - I \cdot R_s \quad (20)$$

The values of the used constants can be found in [24].

In Fig. 6a the simulation results for different voltage sweep rates are depicted. Note that an additional external serial resistance (2.4 kΩ) of the measurement setup is considered in the simulations. Thus, the voltage at the device ($V_{\text{device}}$) is the difference of the applied voltage and the voltage drop at series resistor. This configuration is the reason for the sudden voltage decrease at the device (also called 'snapback') which occurs during SET (step <4> in Fig. 6a). To stay conform to simulations in [24], SET occurs for negative voltages while the RESET voltage is positive.

This model is able to reproduce the measured I-V curve very well. Unfortunately, the model simulation is very sensitive to changes of the input signal, as the authors of [24] have pointed out (compare also [21]). For example, it is not possible to obtain the same shape of the curve when applying a symmetric input signal (±6 V for example). This issue strongly limits applicability of this model for circuit simulations where arbitrary external stimuli may occur. However, the model features highly non-linear switching kinetics, similar to the reported experimental data of [37] for $TiO_2$-based devices (see Fig. 7). This is a major advantage compared to models described in section III.A.

Note that if there is more than one slope in the kinetics plot, different equations are required for modeling each voltage regime; compare [45] for example.

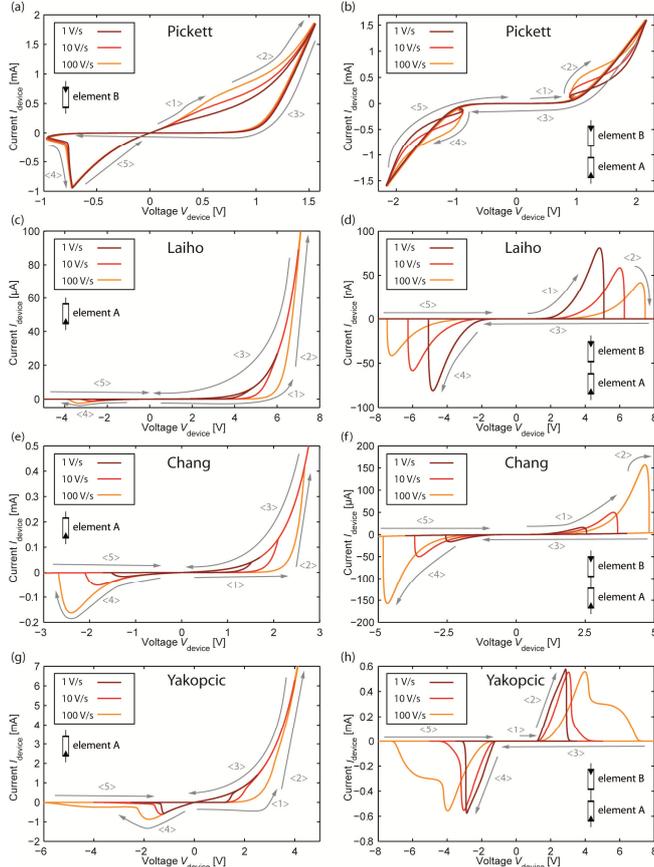

Fig. 6. I-V-characteristics of single as well as complementary cell configuration: simulations for three sweep rates of 1 V/s, 10 V/s and 100 V/s. (a,b) Pickett's model, (c,d) Laiho's model, (e,f) Chang's model and (g,h) Yakopcic's model.

Furthermore, combining two elements anti-serially leads to characteristics offering an ON regime, as required for simulating CRS cells (Fig. 6b). Note that the self-crossing occurring in the ON regime is also observable in Fig. 1b (compare also [33]).

Pickett's and Yang's approach has inspired several other groups to derive generalized models using a hyperbolic sine current relationship offering generic application for memristive devices by simplifying the differential equations.

First, we want to mention two models without any built-in thresholds, introduced by Laiho [26] and Chang [27]. For comparison we also considered a generalized model with built-in thresholds, the model of Yakopcic [29]. SPICE codes as well as applied parameters for all models can be found in [30].

Laiho's model equations are:

$$I = \begin{cases} A_1 \cdot x \cdot \sinh(B_1 \cdot V) & V \geq 0 \\ A_2 \cdot x \cdot \sinh(B_2 \cdot V) & V < 0 \end{cases} \quad (21)$$

$$\dot{x} = \begin{cases} C_1 \cdot \sinh(D_1 \cdot V) \cdot f(x) & V \geq 0 \\ C_2 \cdot \sinh(D_2 \cdot V) \cdot f(x) & V < 0 \end{cases} \quad (22)$$

Here $f(x)$ is the Biolek window function which was added in [30] to limit the range of the state variable to reasonable values. The parameters for this model were selected by the authors with respect to I-V data from [46], an Ag-based ECM device. We simulated this model for three different sweep rates in Fig. 6c and conducted also pulse simulations to obtain the kinetics data (see Fig. 7). The I-V characteristics in Fig. 6c fit well the data in [46]. From Fig. 7, one can see that the extracted kinetics data for this device model offers a relatively weak non-linearity. (Note that the non-linearity of most ECM devices is typically much larger [4].) By combining two elements anti-serially we obtain a characteristic offering very low-conductive ON windows (minimal resistance > 50 MΩ) (Fig. 6d).

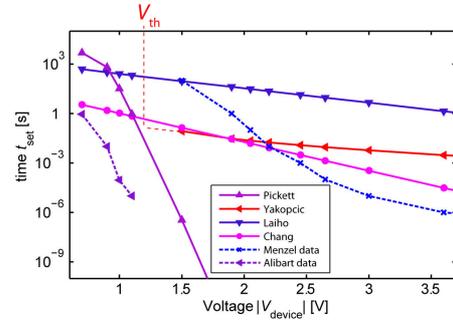

Fig. 7. Set time $t_{SET}$ of the switching from HRS to LRS versus applied pulse height $V_p$. In simulations, the SET time $t_{set}$ is reached when $x$ exceeds $0.5 \cdot (x_{max} + x_{min})$. For Yakopcic's model we extrapolated the curve of $t_{set}$ down to $V_{th}$. (For voltages below $V_{th}$ no switching can occur.)

Curiously, the model predicts a decrease of current level for higher sweep rates – an unexpected behavior for ECM-based CRS cells [47]. Although equations (21)-(22) offer ease of use, predictions from this model should be treated with care. For more details on physics-based modeling of ECM devices we refer to [48, 49]. Note that Menzel's ECM model fulfills all three evaluation criteria, see [48] for criteria one and two, and [47] for anti-serially connected cell simulations.

Chang's model is another generalized sinh model using a more complex set of equations:

$$I = (1-x) \cdot \alpha \left[1 - \exp(\beta V)\right] + x \cdot \gamma \sinh(\delta \cdot V) \quad (23)$$

$$\dot{x} = \lambda \eta_1 \sinh(\eta_2 V). \quad (24)$$

Here, the underlying material system for parameter extraction was a VCM-type $WO_x$ device. The model offers the general trend of higher SET voltages for increasing sweep rates (Fig. 6e). The non-linearity of the kinetics is moderate, in parts similar to data from Menzel for strontium titanate VCM devices, but less steep than Alibart's data (Fig. 7). Moreover, the corresponding anti-serial connection of two devices (Fig. 6f) offers a conceivable behavior for this specific device. An experimental cross-check for our evaluation criteria might verify the model predictions.

Another approach to accommodate highly non-linear kinetics in resistive switches is to reintroduce a built-in threshold. Either current thresholds (e.g., Kvatinsky's TEAM model [28]) or voltage thresholds as in Yakopcic's model [29], which we consider here:

$$I = \begin{cases} a_1 \cdot x \cdot \sinh(b \cdot V) & V \geq 0 \\ a_2 \cdot x \cdot \sinh(b \cdot V) & V < 0 \end{cases} \quad (25)$$

$$\dot{x} = \eta \cdot g(V) \cdot f(x), \quad (26)$$

where $f(x)$ is a window function and $g(V)$ is the function implementing the threshold behavior. Functions $f(x)$ and $g(V)$ read:

$$f(x) = \begin{cases} \exp(-\alpha_p(x - x_p)) \cdot \left(\frac{x_p - x}{1 - x_p} + 1\right) & x \geq x_p \\ \exp(\alpha_n(x + x_n - 1)) \cdot \left(\frac{x}{1 - x_n}\right) & x \leq 1 - x_n \\ 1 & 1 - x_n < x < x_p \end{cases} \quad (27)$$

and

$$g(V) = \begin{cases} A_{pos} \cdot (\exp(V) - \exp(V_{th,pos})) & V > V_{th,pos} \\ -A_{neg} \cdot (\exp(-V) - \exp(V_{th,neg})) & V < -V_{th,neg} \\ 0 & -V_{th,neg} \leq V \leq V_{th,pos} \end{cases} \quad (28)$$

For simulation in Fig. 6g, h the threshold was $V_{th} = V_{th,pos} = V_{th,neg} = 1.2$ V. The I-V behavior in Fig. 6g offers the correct trend, i.e. $V_{SET}$ and $V_{RESET}$ increase for larger sweep rates. For 100 V/s the exponential decay of the window function (equation (27)) determines the RESET process, leading to a gradual OFF switching. In complementary connection this effect is visible for 100 V/s, too. The impact of the assumed voltage threshold becomes also visible in CRS configuration: independent of the applied sweep rate the SET process starts above $V_{th}$. There are two major limitations of this approach one should be aware of. The first one lies in the form of equation (25). As for Laiho's model, if $x \approx 0$ occurs in one device, there is no current flow through the device at all (see Fig. 6h). This property will limit the applicability for simulation of parasitic currents in arrays, for example. Another limitation is inherent for any threshold-based approach where the sub-threshold kinetics is not specified: the device's actual sub-threshold behavior cannot be modeled (see Fig. 7), thus realistic pulse simulations are not possible. An advantage of this model is the flexibility to fit the model to a wide range of I-V curves, i.e. adjust the kinetics in the supra-threshold regime. Note, for the given parameter set, the kinetics is similar to Chang's model (compare Fig. 7).

IV. CONCLUSION

We have analyzed the applicability of existing memristive models for simulation of resistive switches. Our results are:
1) A good way to obtain a realistic dynamic model is to fit I-V characteristic and to incorporate switching kinetics data, as done for Pickett's model or Menzel's ECM model. However, simulation robustness can be an issue for highly complex models as we see for Pickett's model.
2) The derivation of simplified models with better simulation stability is a justifiable first-order approach, as we see for Chang's and Yakopcic's model.
3) Using a generic set of equations which does not directly correspond to the actual real device physics leads to low predictivity, as we see for Laiho's model in complementary configuration.
4) If the basic equations do not reflect the actual device physics well, as we see for the basic memristor equations, with or without window functions, low-predictivity is given at all. However, if only a certain restricted regime of operation shall be modeled also very simple models can be applied.
5) A check for our three evaluation criteria, the I-V characterisistics, the non-linearity of the switching kinetics and the complementary switching behavior of two devices, is a suitable test for model consistency.

Finally, further development of physics-based memristive models is crucial and statements drawn from low-predictive models should be reconsidered carefully.

REFERENCES

[1] ITRS, "The International Technology Roadmap for Semiconductors - ITRS 2011 Edition,", 2011.
[2] J. J. Yang, D. B. Strukov, and D. R. Stewart, "Memristive devices for computing," *Nat. Nanotechnol.*, vol. 8, pp. 13-24, 2013.
[3] M. Di Ventra and Y. V. Pershin, "The parallel approach," *Nat. Phys.*, vol. 9, pp. 200-202, 2013.
[4] I. Valov, R. Waser, J. R. Jameson, and M. N. Kozicki, "Electrochemical metallization memories-fundamentals, applications, prospects," *Nanotechnology*, vol. 22, pp. 254003, 2011.
[5] R. Waser, R. Dittmann, G. Staikov, and K. Szot, "Redox-Based Resistive Switching Memories - Nanoionic Mechanisms, Prospects, and Challenges," *Adv. Mater.*, vol. 21, pp. 2632-2663, 2009.
[6] L.O. Chua and S.M. Kang, "Memristive devices and systems," *Proc. IEEE*, vol. 64, pp. 209-223, 1976.
[7] D. B. Strukov, G. S. Snider, D. R. Stewart, and R. S. Williams, "The missing memristor found," *Nature*, vol. 453, pp. 80-83, 2008.
[8] L.O. Chua, "Memristor-the missing circuit element," *IEEE Trans. Circuit Theory*, vol. CT-18, pp. 507-519, 1971.
[9] L.O. Chua, "Resistance switching memories are memristors," *Appl. Phys. A-Mater. Sci. Process.*, vol. 102, pp. 765-783, 2011.
[10] K. Szot, M. Rogala, W. Speier, Z. Klusek, A. Besmehn, and R. Waser, "TiO$_2$ - a prototypical memristive material," *Nanotechnology*, vol. 22, pp. 254001, 2011.
[11] J. Borghetti, G. S. Snider, P. J. Kuekes, J. J. Yang, D. R. Stewart, and R. S. Williams, "'Memristive' switches enable 'stateful' logic operations via material implication," *Nature*, vol. 464, pp. 873-876, 2010.
[12] H. Kim, M. Pd. Sah, C. Yang, T. Roska, and L. O. Chua, "Memristor Bridge Synapses," *Proceedings of the IEEE*, vol. 100, pp. 2061-2070, 2012.
[13] Y. V. Pershin and M. Di Ventra, "Solving mazes with memristors: A massively parallel approach," *Phys. Rev. E: Stat. Nonlinear Soft Matter Phys.*, vol. 84, pp. 46703, 2011.
[14] F. Corinto, A. Ascoli, and M. Gilli, "Analysis of current–voltage characteristics for memristive elements in pattern recognition systems," *International Journal of Circuit Theory and Applications*, vol. 40, pp. 1277-1320, 2012.
[15] S. Benderli and T. A. Wey, "On SPICE macromodelling of TiO$_2$ memristors," *Electronics Letters*, vol. 45, pp. 377-379, 2009.
[16] Y. N. Joglekar and S. J. Wolf, "The elusive memristor: properties of basic electrical circuits," *Eur. J. Phys.*, vol. 30, pp. 661-675, 2009.
[17] Z. Biolek, D. Biolek, and V. Biolkova, "SPICE Model of Memristor with Nonlinear Dopant Drift," *Radioengineering*, vol. 18, pp. 210-214, 2009.

[18] S. Shin, K. Kim, and S. M. Kang, "Compact Models for Memristors Based on Charge-Flux Constitutive Relationships," *IEEE Trans. Comput-Aided Des. Integr. Circuits Sys*, vol. 29, pp. 590-598, 2010.

[19] T. Prodromakis, B. P. Peh, C. Papavassiliou, and C. Toumazou, "A Versatile Memristor Model With Nonlinear Dopant Kinetics," *IEEE Trans. Electron Devices*, vol. 58, pp. 3099-3105, 2011.

[20] F. Corinto and A. Ascoli, "A Boundary Condition-Based Approach to the Modeling of Memristor Nanostructures," *IEEE Transactions on Circuits and Systems I: Regular Papers*, vol. 59, pp. 2713-2726, 2012.

[21] A. Ascoli, F. Corinto, V. Senger, and R. Tetzlaff, "Memristor Model Comparison," *IEEE Circuits and Systems Magazine*, vol. 13, pp. 89-105, 2013.

[22] R. K. Budhathoki, M. P. Sah, S. P. Adhikari, H. Kim, and L. Chua, "Composite Behavior of Multiple Memristor Circuits," *IEEE Transactions on Circuits and Systems I: Regular Papers*, vol. 60, pp. 2688-2700, 2013.

[23] M. D. Pickett, D. B. Strukov, J. L. Borghetti, J. J. Yang, G. S. Snider, D. R. Stewart, and R. S. Williams, "Switching dynamics in titanium dioxide memristive devices," *J. Appl. Phys.*, vol. 106, pp. 074508, 2009.

[24] H. Abdalla and M. D. Pickett, "SPICE Modeling of Memristors," *IEEE International Symposium on Circuits and Systems (ISCAS)*, pp. 1832-1835, 2011.

[25] R. S. Williams, M. D. Pickett, and J. P. Strachan, "Physics-based memristor models," *IEEE International Symposium on Circuits and Systems (ISCAS)*, pp. 217-220, 2013.

[26] M. Laiho, E. Lehtonen, A. Russel, and P. Dudek, "Memristive synapses are becoming reality," *The Neuromorphic Engineer*, DOI: 10.2417/1201011.003396, 2010.

[27] T. Chang, S.-H. Jo, K.-H. Kim, P. Sheridan, S. Gaba, and W. Lu, "Synaptic behaviors and modeling of a metal oxide memristive device," *Appl. Phys. A - Mater. Sci. Process.*, vol. 102, pp. 857-863, 2011.

[28] S. Kvatinsky, E. G. Friedman, A. Kolodny, and U. C. Weiser, "TEAM: ThrEshold Adaptive Memristor Model," *IEEE Trans. Circuits Syst. I-Regul. Pap.*, vol. 60, pp. 211-221, 2013.

[29] C. Yakopcic, T. M. Taha, G. Subramanyam, and R. E. Pino, "Generalized Memristive Device SPICE Model and its Application in Circuit Design," *IEEE Transactions on Computer-Aided Design of Integrated Circuits and Systems*, vol. 32, pp. 1201-1214, 2013.

[30] C. Yakopcic, T. M. Taha, G. Subramanyam, and R. E. Pino, *Advances in Neuromorphic Memristor Science and Applications - Memristor SPICE Modeling* Springer, 2012, pp. 211-244.

[31] J. Mustafa and R. Waser, "A novel reference scheme for reading passive resistive crossbar memories," *IEEE Trans. Nanotechnol.*, vol. 5, pp. 687-691, 2006.

[32] R. E. Pino, J. W. Bohl, N. McDonald, B. Wysocki, P. Rozwood, K. A. Campbell, A. Oblea, and A. Timilsina, "Compact method for modeling and simulation of memristor devices: Ion conductor chalcogenide-based memristor devices," *IEEE/ACM International Symposium on Nanoscale Architectures*, pp. 1-4, 2010.

[33] S. Schmelzer, E. Linn, U. Böttger, and R. Waser, "Uniform Complementary Resistive Switching in Tantalum Oxide Using Current Sweeps," *IEEE Electron Device Lett.*, vol. 34, pp. 114-116, 2013.

[34] S. Menzel, M. Waters, A. Marchewka, U. Böttger, R: Dittmann, and R. Waser, "Origin of the Ultra-nonlinear Switching Kinetics in Oxide-Based Resistive Switches," *Adv. Funct. Mater.*, vol. 21, pp. 4487-4492, 2011.

[35] Y. Nishi, S. Schmelzer, U. Böttger, and R. Waser, "Weibull Analysis of the Kinetics of Resistive Switches based on Tantalum Oxide Thin Films," *Proceedings of the 43rd European Solid-State Device Research Conference (ESSDERC)*, pp. 174-177, 2013.

[36] S. Yu, Y. Wu, and H. Wong, "Investigating the switching dynamics and multilevel capability of bipolar metal oxide resistive switching memory," *Appl. Phys. Lett.*, vol. 98, pp. 103514, 2011.

[37] F. Alibart, L. Gao, B. D. Hoskins, and D. B. Strukov, "High precision tuning of state for memristive devices by adaptable variation-tolerant algorithm," *Nanotechnology*, vol. 23, pp. 75201, 2012.

[38] E. Linn, R. Rosezin, C. Kügeler, and R. Waser, "Complementary Resistive Switches for Passive Nanocrossbar Memories," *Nat. Mater.*, vol. 9, pp. 403-406, 2010.

[39] M. J. Lee, C. B. Lee, D. Lee, S. R. Lee, M. Chang, J. H. Hur, Y. B. Kim, C. J. Kim, D. H. Seo, S. Seo, U. I. Chung, I. K. Yoo, and K. Kim, "A fast, high-endurance and scalable non-volatile memory device made from asymmetric $Ta_2O_{5-x}/TaO_{2-x}$ bilayer structures," *Nat. Mater.*, vol. 10, pp. 625-630, 2011.

[40] E. Linn, S. Menzel, R. Rosezin, U. Böttger, R. Bruchhaus, and R. Waser, *Nanoelectronic Device Applications Handbook: Modeling of Complementary Resistive Switches* CRC Press, 2013.

[41] D. Biolek, Z. Biolek, and V. Biolkova, "Pinched hysteretic loops of ideal memristors, memcapacitors and meminductors must be 'self-crossing'," *Electronics Letters*, vol. 47, pp. 1385 - 1387, 2011.

[42] E. Linn, S. Menzel, R. Rosezin, U. Böttger, R. Bruchhaus, and R. Waser, "Modeling Complementary Resistive Switches by Nonlinear Memristive Systems," *Proceedings of the 11th IEEE Conference on Nanotechnology*, pp. 1474-1478, 2011.

[43] Y. Ho, G. M. Huang, and P. Li, "Dynamical Properties and Design Analysis for Nonvolatile Memristor Memories," *IEEE Trans. Circuits Syst. I-Regul. Pap.*, vol. 58, pp. 724-736, 2011.

[44] J. J. Yang, M. D. Pickett, X. Li, D. A. A. Ohlberg, D. R. Stewart, and R. S. Williams, "Memristive switching mechanism for metal/oxide/metal nanodevices," *Nat. Nanotechnol.*, vol. 3, pp. 429, 2008.

[45] S. Menzel, S. Tappertzhofen, R. Waser, and I. Valov, "Switching Kinetics of Electrochemical Metallization Memory Cells," *PCCP*, vol. 15, pp. 6945-6952, 2013.

[46] S. H. Jo, T. Chang, I. Ebong, B. B. Bhadviya, P. Mazumder, and W. Lu, "Nanoscale Memristor Device as Synapse in Neuromorphic Systems," *Nano Lett.*, vol. 10, pp. 1297-1301, 2010.

[47] E. Linn, S. Menzel, S. Ferch, and R. Waser, "Compact modeling of CRS devices based on ECM cells for memory, logic and neuromorphic applications," *Nanotechnology*, vol. 24, pp. 384008, 2013.

[48] S. Menzel, U. Böttger, and R. Waser, "Simulation of multilevel switching in electrochemical metallization memory cells," *J. Appl. Phys.*, vol. 111, pp. 014501, 2012.

[49] A. Heittmann and T. G. Noll, "Modeling variability and irreproducibility of nanoelectronic resistive switches for circuit simulation," *18th Asia and South Pacific Design Automation Conference (ASP-DAC)*, pp. 503-508, 2013.